\pdfoutput=1
\documentclass[12pt,preprint]{aastex}
\usepackage{graphicx}
\usepackage{natbib}
\usepackage{aas_macros}
\usepackage[section] {placeins}
\usepackage{subfigure}
\usepackage{float}

\newcommand{\lsim}{\mathrel{\hbox{\rlap{\lower.55ex \hbox{$\sim$}} \kern-.3em \raise.4ex \hbox{$<$}}}}

\def\mout{{\rm\,M_{out}}} 
\def\mgas{{\rm\,M_{gas}}} 
\def\mbh{{\rm\,M_{BH}}} 
\def\mbg{{\rm\,M_{BG}}} 
\def\msun{{\rm\,M_\odot}}

\newcommand{\cc}{\, {\rm cm}^{-3}}

\newcommand{\kms}{\, {\rm km\, s}^{-1}}

\newcommand{\mpc}{\, {\rm Mpc}}
\newcommand{\kpc}{\, {\rm kpc}}
\newcommand{\hmpc}{\, h^{-1} \mpc}

\newcommand{\hi}{\mbox{H\,{\scriptsize I}\ }}
\newcommand{\heii}{\mbox{He\,{\scriptsize II}\ }}

\def\h2{${\rm\,H_2}$}

\makeatletter 

\def\hi{\noindent \hangindent=2.5em}

\def\kpc{{\rm\,kpc}}

\def\mpc{{\rm\,Mpc}}

\def\kms{{\rm\,km/s}}

\def\vol#1  {{{#1}{\rm,}\ }}

\def\hi{{H\thinspace I}\ }

\def\heii{{He\thinspace II}\ }

\newcount\refno
\newcount\rfno
\def\eq{$^{\the\refno\ }$\advance\refno by 1}
\def\ad{\advance\rfno by 1}

\def\clock{\count0=\time \divide\count0 by 60
     \count1=\count0 \multiply\count1 by -60 \advance\count1 by \time
     \number\count0:\ifnum\count1<10{0\number\count1}\else\number\count1\fi}

\def\myputfigure#1#2#3#4#5%
{\hskip0.03\textwidth\vskip#5pt
\makebox[0pt]{\hskip#2in
\includegraphics[width=#3\textwidth]{#1}}\vskip#4pt\hfill}

\newcount\refno
\newcount\rfno
\def\eq{$^{\the\refno\ }$\advance\refno by 1}
\def\ad{\advance\rfno by 1}

\begin{document}

\title{Physics of Coevolution of Galaxies and Supermassive Black Holes}
 
\author{
Renyue Cen$^{1}$
} 
 
\footnotetext[1]{Princeton University Observatory, Princeton, NJ 08544;
 cen@astro.princeton.edu}

\begin{abstract} 

A new model for coevolution of galaxies and supermassive black holes (SMBH) is presented that is physically based.
The evolutionary track starts with an event that triggers a significant starburst 
in the central region of a galaxy.
In this model, the main SMBH growth takes place in post-starburst phase 
fueled by recycled gas from inner bulge stars in a self-regulated fashion on a time scale that 
is substantially longer than $100~$Myrs and at a diminishing Eddington ratio with time.
We argue that the SMBH cannot gorge itself during the starburst phase, despite the abundant supply of cold gas,
because star formation is a preferred mode of gas consumption in such an environment than accretion to the central SMBH.
We also show that feedback from star formation is at least as strong as that from AGN
and thus, if star formation is in need of being quenched, AGN feedback generally does not play the primary role.
The predicted relation between SMBH mass and bulge mass/velocity dispersion is consistent with observations.
A clear prediction is that early-type galaxy hosts of high Eddingtion rate AGNs are
expected to be light-blue to green in optical color, gradually evolving to the red sequences with decreasing AGN luminosity.
A suite of falsifiable predictions and implications with respect to 
relationships between various types of galaxies and AGN, and others, are made.
For those where comparisons to extant observations are possible, the model appears to be in good standing.

\end{abstract}
 
\keywords{Methods: analytic, ISM: kinematics and dynamics,
Galaxies: interactions,
Galaxies: evolution,
Supermassive black holes}
 
\section{Introduction}

The tight correlation between galactic center supermassive black hole (SMBH) mass ($\mbh$) and the bulge
mass ($\mbg$) or velocity dispersion ($\sigma$) in the nearby universe \citep[e.g.,][]{1998Richstone,2000Ferrarese,2002Tremaine}
strongly suggests coevolution of the two classes, at least over the Hubble time.
In many semi-analytic calculations one of the most adopted assumptions, to put it simply, 
is that active galactic nuclei (AGN) feedback 
is able to prevent most of the gas from accreting onto the SMBHs and at the same time
is able to fix most of the ``defects" of galaxy formation models such as the shape 
of the galaxy luminosity function and star formation (SF) history
\citep[e.g.,][]{2000Kauffmann,2006Croton,2008Somerville}
with the underlying feedback physics parameterized.
The substantial success in explaining a variety of observations enjoyed
by these semi-analytic models is indicative of the relevance of AGN feedback. 
Calculations of the coupled evolution of SMBHs and galaxies using three-dimensional hydrodynamic simulations 
deploy thermal energy feedback in regions significantly outside of 
the Bondi radius of the putative SMBH that effectively couples to the surroundings 
to regulate the SF and eventually drive the gas away.
These pioneering detailed simulations have provided much physical insight
and appear to be remarkably successful in accounting for many intricate 
observables, including AGN light curves, Eddington ratio distributions and
SMBH-bulge relation and its scatter, for certain chosen value of the feedback energy strength \citep[e.g.,][]{2005DiMatter,2006Hopkins}.
What is hitherto left open in these calculations is the physical origin of the adopted energy feedback.
One concern is that the derived SMBH-bulge relation depends very sensitively on the adopted energy feedback parameter
due to the strong radiative cooling \citep[e.g.,][]{2010Silk, 2011Choi}.
Thus, it is prudent to seek underlying physical origins for these successful models and,
before that is achieved, continue to explore alternative models.

This paper synthesizes an alternative physical model largely based on known physics.
Before describing our overall model, we shall first,
in \S 2, examine the plausibility of the fundamental claim
that AGN feedback is primarily responsible for regulating not only SMBH growth but also SF.
We argue that scenarios invoking AGN as the primary ``blowing machine" 
during the intense starburst phase
may logically require significant fine-tuning.
We then describe the evolutionary path from a starburst to an elliptical galaxy,
including the coupled evolution of star formation and SMBH growth in the ensuing two sections.

In \S 3, we show that growth of SMBH during the starburst phase 
is limited and constitutes a small fraction of the overall SMBH consumption.
The physical reason is that this phase is over-supplied with gas such that
only a very small central
disc is gravitationally stable (Toomre parameter $Q>1$) for gas accretion onto the SMBH, 
while all other regions are unstable and more conducive to star formation.
Since the SF time scale 
is much shorter than the Salpeter accretion time scale, 
most of the gas forms into stars.
The accreted mass during this phase is probably limited to a few percent of the final SMBH mass.
In \S 4, we point out that energy or momentum feedback from 
SF is at least as competitive as that from the AGN during the starburst phase. 
Therefore, SF is largely responsible for blowing most of the last patch of  
gas away to end the starburst phase. 
In short, during the starburst phase, the SMBH does not grow significantly 
and does not play the leading role in quenching the star formation.

In \S 5, we show that most of the growth of the SMBH occurs in the 
ensuing post-starburst period,
when the bulge/elliptical galaxy is largely in place and SF
enters ``passive" evolution.
The fuel for this primary growth phase is provided
by the gas recycled back into the interstellar medium (ISM) from aging bulge stars, 
proposed earlier by \citet[][]{1988Norman} in the context of a central stellar cluster
and stressed recently by \citet[][]{2007Ciotti} in the context of elliptical galaxies.
It provides a relatively ``diffuse" (compared to the starburst phase) but steady gas supply 
that, we show, is ideal for feeding SMBH 
via an accretion disc.
Meanwhile, SF is the dominant mode for gas consumption in the outer region 
because the accretion is unstable to fragmentation there, even in this phase.
Self-regulation is at work for the growth of the SMBH during this period and is provided
by much more robust (compared to energy feedback) radiation pressure induced momentum.
The amplitude and slope of the resultant SMBH-bulge relation 
with this self-regulation is consistent with observations.

In this model,
the entire evolution from the onset of starburst, due to a gas-rich merger or some  
significant event that drives a large amount of gas into the central region within a short period of time,
to becoming a quiescent elliptical galaxy (or a bulge of a future spiral galaxy)
consists of three distinct periods, as summarized in \S 5.1 and in Figure \ref{fig:evol}:
(1) ``Starburst Period": merger of two gas-rich spiral galaxies or 
some other significant event induces a starburst that lasts about $10^7-10^8$yrs.
The SMBH grows modestly during this period.
The feedback energy/momentum from the starburst, i.e., supernovae,
drives the last patch of gas away and helps shut down star formation.
(2) ``SMBH Prime Period": several hundred million years 
after the end of the starburst, aging low-to-intermediate mass stars,
now in the form of red giants and other post-main-sequence states,
start to return a substantial fraction of their stellar mass to the ISM.
The SMBH accretion is mostly supply limited in most of this period, except during the first several
hundred million years or so, and lasts for order of gigayear.
Because the rate of gas return from stars diminishes with time,
the Eddington ratio of the SMBH decreases with time and
the SMBH spends most of the time during this period at low  
the Eddington ratio ($\le 10^{-3}$).
The SMBH growth is nearly synchronous with star formation from recycled gas during this period.
The accompanying star formation rate is quite substantial, roughly
$\sim (5-10) (M_*/10^{11}\msun)(t/1{\rm Gyr})^{-1.3}\msun$~yr$^{-1}$,
where $t$ is time in Gyr and $M_*$ is stellar mass of the elliptical galaxy formed
during the starburst (at $t=0$).
The duration of this phase {\it depends sensitively} on the lower cutoff mass of the initial mass function (IMF).
(3) ``Quiescent Elliptical Galaxy": several gigayears after the end of the starburst
the elliptical galaxy is now truly red and dead - gas return rate is now negligible
so both accretion to the central SMBH and residual star formation have ceased.
It is possible, at least for an elliptical galaxy that is not too massive
(i.e., $M_{\rm tot}\le 10^{12}\msun$), that it may grow a disk.
The feeding of the central SMBH in the bulge of spiral galaxy during this period is no longer by aging stars,
rather by occasional objects (molecular clouds, stars, etc) 
that happen to be on some plunging orbits 
due to secular or random events.

We present some predictions and implications
of this model in \S 6.2-6.9, followed by conclusions in \S 7.
Where comparisons can be made between the predictions of the model and 
observations, they appear to be in good agreement.
Some additional predictions could provide further tests of the model.

\section{AGN Cannot Regulate Star Formation During Starburst}

While the subsequent sections of quantitative physical analysis
are independent of statements made in this section,
we shall argue for the assertion in the title of this section 
with logic, in hopes of being able to provide some conceptual clarity to
the role of AGN feedback on star formation during the starburst phase.
The starting point of the evolutionary sequence is a starburst.
It may be triggered by a major merger of two gas-rich galaxies
or by other significant events that channel
a large amount of gas into the central region in a short period of time.
Consider that an event causes a large amount of gas of mass $\mgas$
to land in the central region. 
Physical processes then operate on the gas to produce a starburst,
accompanied by some growth of the central SMBH, along with some associated feedback from both.
Extreme events of this kind may be identified with observed 
Ultra-Luminous InfraRed Galaxies (ULIRGs) \citep[e.g.,][]{1988Sanders}
or Sub-Millimeter Galaxies (SMGs) \citep[e.g.,][]{2005Chapman}.
Theoretical models \citep[e.g,][]{1998Silk, 2006Hopkins}
have proposed that feedback from AGN 
is responsible for the regulation of SF and SMBH growth so as to produce 
the observed \citet[][]{1998Magorrian} 
relation where the ratio of the final SMBH to bulge stellar mass
is $\mbh:\mbg\sim 2:1000$.
We shall now re-examine this case.

Consider how the infallen gas may be partitioned.
Mass conservation requires $\mbh+\mbg+\mout=\mgas$,
where $\mout$ is the amount of gas that is blown away from the bulge.
Clearly, only a very small fraction of
the initial gas $\mgas$ can possibly end up in the central SMBH, i.e., $f_{\rm BH}\equiv \mbh/\mgas\ll 1$.
Let us assume that the reason for a very small $f_{\rm BH}$ is because
the feedback from the central SMBH prevented its own further growth during this phase.
Since SMBH masses are observed to span a very wide range, it must be
that this purported SMBH feedback process that regulates its own growth 
is galaxy specific, i.e., dependent on at least some physical variables
characterizing the galaxy.
A usual and reasonable assumption (which we are not advocating at the moment) 
for that is that either the gravitational potential well of the bulge or of the total halo determines
the final SMBH mass, in coordination with its feedback.

Does SMBH feedback dominate that 
of starburst in terms of regulating both SMBH growth and starburst?
While we will show later (in \S 3) that the answer is largely no to regulating the starburst at least,
we assume that the answer is yes to both for the sake of continuing the present thought experiment.
The simplified sequence of events then plays out more or less as follows.
The central SMBH accretes gas and builds up its feedback strength 
until its mass has reached the observed value,
then blows away all the remaining gas and both SMBH accretion and SF stop abruptly.
What might have happened to SF during all this time before
the gas is blown away?
There are three possible scenarios.
Scenario \#1, the SMBH accretion is so competitive and quick that most of the gas 
is blown away by the SMBH feedback before much SF has occured.
That of course cannot have happened, because that would be inconsistent with the observed $\mbh-\mbg$ relation.

Scenario \#2, SF precedes at a pace that is in concert with the SMBH feedback
such that by the time that $\mbh=0.002 (\mgas-\mout)$,
the amount of stars formed is equal to $\mbg=0.998 (\mgas-\mout)$;
the rest of gas of mass $\mout$ got blown away by the feedback from the SMBH.
This scenario is designed to match the observed $\mbh-\mbg$ relation.
What remains undetermined is how large $f_{\rm out}\equiv \mout/\mgas$ is.
Is it close to $1$ or $0$?
In the case $f_{\rm out}\sim 1$,
because $(1-f_{\rm out})$ is a small number, there is no particular 
preferred value for it.
The potential well created by the eventual bulge stars would be much shallower
than the original one already created by the residing gas. 
In other words, the SMBH only knew the potential well of the original gas and 
it would be rather arbitrary how much stars the SMBH decides to allow the bulge to have.
If one argues that it is the potential well of the total halo mass
that matters,
the SMBH still did not know how to let SF take place at such a rate that
we have the very tight observed $\mbh-\mbg$ relation for the {\it bulge region}.
Thus, this case also appears to require much fine tuning.
Besides, if $(1-f_{\rm out})$ is too small, the bulge will be too small compared to what is observed.

The opposite case with $f_{\rm out}\ll 1$ is at least substantially more stable,
since a large fraction 
of the original gas has formed into stars 
before the remainder of the gas got blown away.
In this case the SMBH would ``know better"
the gravitational potential well eventually sustained by bulge stars,
because it is not too far from that created by the initial gas.
Then, how did the SMBH know when to blow away the remaining gas left over
from SF and SMBH accretion?
Should the SMBH blow away the gas when $f_{\rm out}=0.90$ (an arbitrarily picked number for illustration purpose)
or should it wait a bit longer to finally blow away the gas 
when $f_{\rm out}=0.10$?
It may require more energy or momentum in the former 
than the latter;
but that can readily be accommodated by a proportionally increased amount of gas accreted,
in the vein of feedback from SMBH providing the required feedback energy or momentum.
Since the amount of gas available before $f_{\rm out}=0.90$ is blown away in this hypothetical case
is capable of growing the SMBH to be 900 more massive than observed and
the amount of time available (cosmological scale) is much longer than Salpeter time,
there is no obvious reason why the SMBH cannot grow 10 times (or whatever factor) larger to 
blow away the gas when $f_{\rm out}=0.90$ instead of when $f_{\rm out}=0.10$.
How the SMBH has communicated with the bulge to ration the gas consumption would be a mystery.
Thus, even in this case with $f_{\rm out}\ll 1$,
taking it as a given that the SMBH always stands ready to provide the necessary feedback,
having SMBH feedback to regulate the overall SF in the bulge
such that the ratio of the two matches the observation, again, requires a substantial amount of fine tuning.
Nevertheless, since it is reasonable to expect that 
the dependence of the outcome, such as the $\mbh-\mbg$ relation,
on any proposed feedback processes
(including those based on thermal energy deposition near the galaxy center)
is likely a monotonic function of the adopted feedback strength, 
it should be expected that 
a solution be found such that the observed $\mbh-\mbg$ relation is obtained, for some chosen value of 
feedback strength, at least for some narrow range in $\mbg$.
But, until there is clear physical reason or direct observational evidence 
to support the chosen value of the feedback parameter which
the solution sensitively depends on, such an approach remains to be refined.
We will provide an alternative, significantly less contrived, quantitative physical mechanism to circumvent this
concern of fine tuning.

\section{Starburst Phase: Modest SMBH Growth and SF Shutdown by Stars}

We have argued in the previous section that AGN feedback cannot logically play the leading role in regulating SF,
in the sense that while some feedback from the SMBH can certainly affect its surrounding gas,
there is no particular reason why this could provide a quite precise (within a factor of a few) rationing mechanism
during the starburst phase so as to produce the observed relation between the two.
We shall now argue for Scenario \#3: 
{\it during the starburst phase} the SF is self-regulated and self-limited, 
while SMBH growth is modest, does not need regulation and does not provide significant feedback to star formation.

We now give a physical reason for why, even though there is a very large supply of gas 
in the bulge region during the starburst phase,
the SMBH growth is modest.
We will make three simplifying assumptions to present trackable 
illustration without loss of generality. 
We assume (1) for the regions of interest a geometrically thin Keplerian disc dominated by the SMBH gravity (at least at the radii of interest here)
is in a steady state, meaning the accretion rate \citep[][]{1992Frank}:
\begin{equation}
\label{eq:Mdot}
\dot M = 3\pi \nu \Sigma_g \left[1-(r_{\rm in}/r)^{1/2}\right]^{-1} \approx 3\pi \nu \Sigma_g 
\end{equation}
\noindent
is constant in radius $r$ and time, where $\Sigma_g$ is gas mass surface density and $\nu$ is viscocity;
the last equality is valid because the radii of interest here are much larger than the radius of the inner disc $r_{\rm in}$;
note that it is inevitable to form a disk in the central given the rapid cooling and finite angular momentum;
(2) we adopt the $\alpha$-disc viscosity \citep[][]{1973Shakura}:
\begin{equation}
\label{eq:nu}
\nu = \alpha c_s^2 \Omega^{-1},
\end{equation}
\noindent
where $\alpha$ is a dimensionless viscosity constant 
for which magnetorotational instability process \citep[][]{1991Balbus}
provides a physical and magnitude-wise relevant value;
$c_s$ is sound speed and $\Omega$ is angular velocity 
(equal to epicyclic frequency for Keplerian disc).
The Toomre Q parameter of the gas disc can be obtained from Equations (\ref{eq:Mdot},\ref{eq:nu}): 
\begin{eqnarray}
\label{eq:Q}
Q\equiv {c_s\Omega\over \pi G\Sigma_g} = {1\over 3^{1/2}\pi^{3/2}\alpha^{1/2}}\left({\dot M\over \mbh}\right)^{1/2}\left({G^{-1/4}\mbh^{5/4}\over \Sigma_g^{3/2} r^{9/4}}\right) 
\end{eqnarray}
\noindent
where $G$ is gravitational constant.
The slope of the surface brightness profiles of the inner region of the observed powerlaw elliptical galaxies,
which are assumed to be the product of the starbursts resulting from the gas-rich galaxy mergers,
has a value concentrated in the range $-1.0$ to $-0.5$ \citep[e.g.,][]{1997Faber,2009Kormendy},
reproduced in merger simulations \citep[e.g.,][]{2009Hopkins}.
Presumably the initial gas density profile is similar to the final observed stellar density profile
in the inner regions.
For ease of algebraic manipulations, 
we assume (3) the de Vaucouleur mass surface density profile (with a half-mass radius $r_e$) but with the
inner region at $r\le r_p \equiv 0.07r_e$ modified to be a Mestel disc as: 
\begin{equation}
\label{eq:Sigma}
\Sigma_g(r)= \Sigma_0 \left({r\over r_0}\right)^{-1}\quad\quad {\rm for}\quad\quad r\le r_p,
\end{equation}
\noindent
where $\Sigma_0$ is the normalizing surface density at some radius $r_0$;
we will only be dealing with region $r\le r_p$;
the notional nuclear velocity dispersion of the system without the central SMBH at $r\le r_p$ 
is related to $\Sigma_0$ and $r_0$ by
\begin{equation}
\label{eq:sigma2}
\sigma_n^2= \pi G \Sigma_0 r_0.
\end{equation}
\noindent
Subsequent results do not sensitively depend on the exact slope.
The total mass of such a hybrid profile is equivalent to a truncated isothermal sphere 
with a truncation radius of $2r_e$ and velocity dispersion on galactic scales of $\sigma_g$ such that 
\begin{equation}
\label{eq:sigmasigma}
\sigma_n= 1.55 \sigma_g.
\end{equation}
\noindent
Since the dynamical time, say at $1$kpc for a $200\kms$ bulge being only
$5\times 10^6$yr,  is much shorter than the Salpeter time,
it is appropriate to assume that the gas disc is assembled instantaneously with respect to
accretion to the SMBH when infalling gas lands on the disc.
Combining Equations (\ref{eq:Q},\ref{eq:Sigma},\ref{eq:sigma2},\ref{eq:sigmasigma}) we rewrite $Q$ as
\begin{eqnarray}
\label{eq:Q2}
Q = 0.32 \alpha_{0.01}^{-1/2} \epsilon_{0.1}^{-1/2}l_E^{1/2}M_8^{5/4}\sigma_{200}^{-3}r_{\rm pc}^{-3/4}, 
\end{eqnarray}
\noindent
where $\alpha_{0.01}=\alpha/0.01$, 
$\epsilon=0.1\epsilon_{0.1}$ is the SMBH radiative efficiency,
$l_E$ is Eddington ratio,
$M_8=\mbh/10^8\msun$,
$\sigma_{200}=\sigma_g/200\kms$,
$r_{\rm pc}=r/1{\rm pc}$.
The value of $\alpha$ is quite uncertain, possibly ranging from $10^{-4}$ to $1$
\citep[e.g.,][]{1995Hawley, 1995Brandenburg, 1996Stone, 1998Armitage, 2001Gammie, 2003Fleming,2007Fromang}.
Setting $Q$ in Equation (\ref{eq:Q2}) to unity defines the disc stability radius 
\begin{equation}
\label{eq:rQ}
r_Q = 0.22 \alpha_{0.01}^{-2/3} \epsilon_{0.1}^{-2/3}l_E^{2/3}M_8^{5/3}\sigma_{200}^{-4}~{\rm pc} 
\end{equation}
\noindent
within which $Q>1$ and disc is stable to gravitational fragmentation,
and outside which $Q<1$ and disc is subject to gravitational fragmentation to form stars,
supported by both simulations \citep[e.g.,][]{2001Gammie,2003Rice}
and circumstantial observational evidence of the existence of stellar disc at small Galactic radius ($\sim 0.1$pc) \citep[e.g.,][]{2003Levin,2006Paumard}.
The demarcation value of $Q$ between stability and fragmentation does not appear to be qualitatively different even
if the disc is under strong illumination \citep[e.g.,][]{2003Johnson},
as might happen to a nuclear gas disc in the starburst phase.
The disc mass within $r_Q$ is 
\begin{eqnarray}
\label{eq:MQ}
M_Q = 9.8\times 10^6 \alpha_{0.01}^{-2/3} \epsilon_{0.1}^{-2/3}l_E^{2/3}M_8^{5/3}\sigma_{200}^{-2}\msun  
\end{eqnarray}
\noindent
{\it This is the accretable mass out of the entire bulge region} 
(note that some of the outer regions are more random motion supported).
This conclusion reached is in good agreement with 
\citet[][]{2003Goodman}, who employs somewhat different assumptions
than in this study in that he assumes local energy balance,
while we impose the observationally inferred inner density profile to be self-consistent;
the good agreement suggests that this result is quite robust, insensitive to assumptions made.
Taking cue from our own Galaxy, if we assume that the initial SMBH mass 
of the two merging spiral galaxies of mass $\sim 10^{12}\msun$ each is $2.5\times 10^6\msun$,
for a spiral galaxy of velocity dispersion of $200\kms$,
we see that the amount of mass that could be readily accreted according to 
Equation (\ref{eq:MQ}) 
using $M_8=0.05$ 
is $6.7\times 10^4 \alpha_{0.01}^{-2/3}\epsilon_{0.1}^{-2/3}l_E^{2/3}\msun$.
Note that the final SMBH mass for such a system is $\sim 1.3\times 10^8\msun$ \citep[][]{2002Tremaine},
if we were to match the observations.

It is possible that the mass accreted to the SMBH may be larger than that
indicated by Equation (\ref{eq:MQ}) due to replenishment.
Replenishment of low angular momentum gas during the starburst phase
may be possible in two ways: (1) through orbital decay of outer disc gas
or (2) direct infall of low-J gas from outer regions.
We will show below that (1) does not significantly increase the accretable mass.
Process (2) is probably unavoidable to some extent 
but unlikely to be frequent enough to be significant for the following reasons.
All the low angular momentum infalling gas falls into the inner regions initially according to its 
respective specific angular momentum driven by the torque of the trigger event
(e.g., merger or some other significant torquing event).
To replenish low angular momentum gas directly to the central region
some frequent and significant torquing events during the starburst phase are needed.
It seems unlikely that such events will be frequent enough to
be able to reach the observed final SMBH mass:
about $\sim 100-1000$ replenishments will be required.
One might approximately equate the number of replenishment (i.e., significant disturbance)
to the number of generations of stars formed during the starburst phase (by assuming that each generation of star formation
manages to redistribute the angular momentum of a significant fraction of the gas), 
which is unlikely to be close to $\sim 100-1000$.
In summary, taking into account possible additional accretion due to some replenishment
and giving the benefit of the possibility of $\alpha_{0.01}<1$, 
it seems improbable that the SMBH is able to acquire a mass during the starburst phase
that would be much more than $10\%$ of the final value.

At $r\ge r_Q$, the disc is unstable to SF. 
For SF under the conditions relevant here
both the dynamical and cooling time are short and do not constitute significant bottleneck; 
if they were the only time scale bottleneck, SF would be too efficient.
A possible bottleneck for SF is the time scale to rid the cloud of the magnetic flux
(assuming the SF clouds are initially magnetically sub-critical).
The main ionization source in the depth of molecular cloud cores is cosmic rays (CR).
While the exact ionization rate by CR is unknown for other cosmic systems,
we have some estimate of that for our own Galaxy,
$\zeta_{\rm CR,Gal}=(2.6\pm 1.8)\times 10^{-17}$~s$^{-1}$
\citep[e.g.,][]{2000vanderTak}.
If one assumes that the CR ionization rate in starburst is $100$ times (modeling a typical ULIRG in this case)
that of the Galactic value, considering that the SF rate
in ULIRGs is $100-1000$ times the Galactic value occuring in a more compact region and 
that the CR in ULIRGs may be advected out via fast galactic winds (versus slow diffusion in the Galaxy),
one may roughly estimate that the ambipolar diffusion time is $7\times 10^6$yr at a density of $n\sim 10^5\cc$
using standard formulas for recombinations \citep[e.g.,][]{2007McKee}.
This estimate is, however, uncertain.
We will again look to direct observations to have a better gauge.
\citet[][]{2004Gao} show, from HCN observations,
that ULIRGs and LIRGs convert molecular gas at $n\ge 3\times 10^4\cc$ 
at an e-folding time scale of $t_{\rm SF}\sim 2$Myr,
consistent with the above rough estimate.
It is clear that SF time scale is much shorter than the Salpeter time of $4.5\times 10^7\epsilon_{0.1}$yr;
in other words, when gas is dense and unstable,
star formation competes favorably with the SMBH accretion with respect to gas consumption. 
Therefore, most of the gas at $r\ge r_Q$ will be depleted by SF.
When the density profile of the disc at $r\ge r_Q$ steepens to be $\Sigma_g(r) = \Sigma_Q (r/r_Q)^{-5/2}$,
where $\Sigma_Q$ is the gas surface density at $\sim r_Q$,
the disc at $r\ge r_Q$ may become stable again.
While continued accretion supplied by gas on the outer disc is likely, albeit at a much lower level,
the mass integral is convergent and most of the mass of this outer disc is at $r_Q$ given the density slope,
even if the entire outer disc at this time is accreted.

Thus, it appears that the amount of gas that is actually accreted by the SMBH during the starburst phase
is rather limited.
This new and perhaps somewhat counter-intuitive 
conclusion is strongly supported by available observations of ULIRGs.
This conclusion is also opposite to most models that rely on SMBH to provide
the necessary feedback to regulate star formation
\citep[e.g,][]{1998Silk, 2006Hopkins}.
Observational evidence is that the SMBHs in ULIRGs and SMGs
appear to be significantly smaller (an order of magnitude or more) than what 
the $\mbh-\mbg$ relation would suggest 
\citep[e.g.,][]{1998Genzel,2000Ivison,2003Ptak,2004Ivison, 2005Alexander,2005bAlexander,2006Kawakatu, 2008Alexander}.
Nonetheless, it is expected that the AGN contribution in ULIRGs 
should become relatively more important for 
larger more luminous galaxies (see Equation \ref{eq:MQ}), consistent with observations \citep[e.g.,][]{1998Lutz}.
Starbursts occuring on rotating nuclear disc/rings in ULIRGs are also supported
by circumstantial observational evidence \citep[e.g.,][]{1998Downes}.

The overall conclusion that the SMBH feedback has little effect on the amount of stars
formed is in agreement with that of \citet[][]{2010Debuhr} who investigated
the radiation pressure-regulated SMBH feedback in {\it the starburst phase} of the merger simulations
utilizing a sub-grid model for SMBH accretion. 
One specific common outcome between our calculation and their simulation
is that most of the gas formed into stars, regardless of the feedback strength.
A notable difference between our calculation and theirs 
is that their simulation resolution, a gravitational softening length of $47~$pc, 
is significantly larger than $r_Q$
(Equation \ref{eq:rQ}). As a result, it is possible that their simulations do not 
resolve small scale that separates stable accretion
from unstable, fragmenting disc, which is crucial to our quantitative conclusion
(note that they use the viscosity parameter $\alpha=0.05-0.15$ that is larger than our fiducial value of $0.01$, which
would yield a still smaller $r_Q$, see Equation \ref{eq:MQ}).
Thanks to that difference, we were able to conclude that, even without considering any feedback from the central SMBH,
the SMBH during the starburst phase
does not grow to anywhere close the observed final mass,
because star formation can more favorably deplete the gas that may otherwise accrete to the SMBH,
whereas they find SMBH masses to be too large even with substantial feedback
(note that they use 10 times $L/c$ radiation pressure force assuming multiple scatterings 
of each converted FIR photon).
It seems likely that their different conclusion may be due to a much higher accretion rate at
their resolution scale, which we argue does not reflect the actual accretion onto the SMBH,
but rather the disc is unstable at that scale and mostly forms stars.
As we have noted in the previous paragraph,
observations indicate that the SMBH masses in the starburst phase
appear to be smaller than the final values seen in quiescent elliptical galaxies
by an order of magnitude, consistent with our conclusion.
Substantially higher resolution (a factor of $\sim 100$) simulations may be necessary 
in order to realistically and more accurately simulate the intricate 
competition between accretion and star formation.

\section{A Comparison of Feedback Energetics Between Star Formation and SMBH}

Having shown the unlikelihood of substantially growing SMBH during the starburst phase,
we now turn to a comparison of the energetics of SMBH and SF to show that,
feedback from starburst itself should play the leading role
in shutting down or quenching star formation, i.e., promptly sweeping away the 
final portion of the gas, where needed.

\begin{deluxetable}{llll}
\tablecolumns{4}
\tablewidth{0pc}
\tablehead{
\colhead{\#} & \colhead{Form} & \colhead{SF}  & \colhead{SMBH}}
\startdata
(1) & total radiation & $\epsilon_*({\rm rad})=7\times 10^{-3}$ & $\epsilon_{\rm BH}({\rm rad})=2\times 10^{-4}$   \\
(2) & ionizing radiation ($\ge 13.6$eV) & $\epsilon_*({\rm LL})=1.4\times 10^{-4}$ & $\epsilon_{\rm BH}({\rm LL})=3\times 10^{-5}$   \\
(3) & X-ray ($2-10$keV) & $\epsilon_*($2-10${\rm keV})=9\times 10^{-8}$ & $\epsilon_{\rm BH}($2-10${\rm keV})=5\times 10^{-6}$   \\
(4) & mechanical & $\epsilon_*({\rm SN})=1\times 10^{-5}$ & $\epsilon_{\rm BH}({\rm BAL})=(0.2-2.8)\times 10^{-5}$   \\
(5) & radio jets & $\epsilon_*({\rm jet})=0$ & $\epsilon_{\rm BH}({\rm jet})=4\times 10^{-5}$   
\enddata
\tablecaption{Comparison of SF and SMBH energetics:
Under the assumption that $\mbh:\mbg=2:1000$, a Salpeter IMF for stars and a radiative efficiency of SMBH accretion of $10\%$ 
\citep[][]{2002Yu}, 
energy output from both SF and SMBH in various forms are listed:
(1) total radiation energy,
(2) ionizing radiation,
(3) X-ray radiation in $2-10$keV band,
(4) mechanical energy and (5) radio jets.
\label{table1}}
\end{deluxetable}

To avoid any apparent bias against SMBH or a possibly circular looking argument by the assertion 
that most of the SMBH growth takes place in the post-starburst phase (as we will show in \S 4),
we shall for the moment generously assume that 
the entire SMBH growth occurs during the starburst phase,
to maximize the energy output from the SMBH,
when comparing the energetics from the SMBH and the starburst.
In Table 1, under the assumptions that $\mbh:\mbg=2:1000$, a Salpeter IMF for stars and a radiative efficiency of SMBH accretion of $10\%$, 
energy output from both SF and SMBH in various forms are listed:
(1) total radiation energy,
(2) ionizing radiation,
(3) X-ray radiation in $2-10$keV band,
(4) mechanical energy, which is supernova explosion energy for SF and broad absorption line (BAL) outflow for SMBH, respectively,
and (5) radio jets.
To obtain energy is ergs per $M_{\rm star}$ formed, one just needs multiply
each coefficient in Table 1 by $M_{\rm star}c^2$, where $c$ is speed of light.
The relevant references are
\citet[][]{1994Elvis} and \citet[][]{2004Sazonov} 
for both $\epsilon_{\rm BH}(LL)$ and $\epsilon_{\rm BH}($2-10${\rm keV})$,
\citet[][]{2003Ranalli} for $\epsilon_*($2-10${\rm keV})$,
\citet[][]{2009Moe} and \citet[][]{2010Dunn} for $\epsilon_{\rm BH}({\rm BAL})$
and \citet[][]{2006Allen} for $\epsilon_{\rm BH}({\rm jet})$ (if one uses energy seen
in the most powerful radio jet lobes and assumes that they are produced by the most massive SMBH,
a comparable value is obtained).
The entry for the BAL energy is based on two cases and very uncertain,
primarily due to lack of strong constraints on the location of the BAL and
their covering factor.

It is evident that aside from the energy in the form of radio jets and hard X-rays,
SF is at least competitive compared to SMBH.
Heating due to hard X-rays from SMBH via metal line or Compton heating 
affects only the very central region surrounding the SMBH, not over the entire galaxy \citep[][]{2007Ciotti}.
Within the physical framework outlined here,
most of the SMBH growth occurs post-starburst 
and radio jets occur at a still later stage in core elliptical galaxies,
energy output (or momentum output derived from it) from SF in all relevant forms
should dominate over that of SMBH.
Our argument that radio jets occur at a later stage in galaxy evolution
is not at present based on a physical model, but on empirical evidence.
Observationally, it appears that all significant radio jets 
are launched in elliptical galaxies that have flat cores \citep[][]{2006Balmaverde},
with a very few exception that originated in  
disc galaxies \citep[e.g.,][]{1999Evans,2001Ledlow} or S0's \citep[e.g.,][]{2001VeronCetty}.
But none has been associated with elliptical galaxies with an inner powerlaw brightness profile slope.
It has been plausibly argued that
powerlaw elliptical galaxies are produced by gas-rich mergers (we adopt this scenario
where a powerlaw elliptical galaxy is produced following each major gas-rich merger triggered starburst)
\citep[e.g.,][]{1997Faber},
whereas core elliptical galaxies are produced later by dry mergers of two elliptical galaxies
where the flat core is carved out by the merger of the two SMBHs via dynamical friction
\citep[e.g.,][]{2001Milosavljevic}.
Directly supporting this statement is the lack of radio jet 
in available observations of ULIRGs \citep[e.g.,][]{2010Alexander},
in agreement with other observations that indicate a significant time-delay
between starburst and radio activities \citep[e.g.,][]{2006Emonts}.
An independent, additional argument comes from the fact that radio jets are highly collimated 
and, for the most powerful ones that are energetically relevant,
they appear to dissipate most of the energy at scales larger than that of the bulge region,
suggesting that, even if one were to ignore the previous timing argument,
the efficiency of heating by radio jets for the bulge region is likely low and at best non-uniform.
Weaker radio feedback, observed almost exclusively in galaxies with an atmosphere of hot gas,
may be able to steadily provide feedback energy but it is too weak to be energetically important.
Besides, they appear to only operate in elliptical galaxies with hot atmospheres \citep[e.g.,][]{2005Best}.

The amount of supernova explosion energy that couples to the surrounding medium 
is $E_{\rm SN}=1\times 10^{-5}M_*c^2$, which is exactly equivalent to $5\times 10^{-3} \mbh c^2$
used in the influential simulations of \citet[][]{2006Hopkins} with thermal AGN feedback, assuming $\mbh:\mbg=2:1000$.
Because the energy output from supernovae is subject to less cooling than that from
the AGN, since the former is at larger radii and lower densities than
the latter, we expect that the amount of energy due to supernovae 
can at least as effectively as that proposed from AGN
to drive the gas away.
Thus, when most of the gas have formed into stars
(i.e., the bulge is largely in place after $\sim 10^7-10^8$yr of starburst),
the remaining gas should be blown away by collective supernova explosions and the starburst comes to a full stop,
reminiscent of what is seen in the simulations of \citet[][]{2006Hopkins} with AGN feedback.
Detailed high-resolution simulations will be necessary, 
taking into account cooling and other physical processes, 
to ascertain the fraction of gas that is blown away.
In short, the bulk of galactic winds is likely driven by stellar feedback from the starburst.
Galactic winds are observed and casual connection between SF rate and 
wind fluxes has been firmly established  
\citep[e.g.,][]{2001Heckman,2009Weiner}, lending strong observational support for the argument.

\section{Post-Starburst: Main Growth of SMBH with Self-Regulation}

The previous section ends when the starburst has swept away the remaining gas
and ended itself.
This section describes what happens next - the post-starburst period,
the initial period of which is also known as K+A galaxies.

The newly minted (future) bulge enters its ``passive" evolutionary phase,
as normally referred to.
We would like to show that this is when most of the action for SMBH begins,
fueled by recycled gas from aging low-to-intermediate mass stars.
Since two-body relaxation time is much longer than the Hubble time,
it is safe to assume that the stars formed in the inner region 
during the starburst phase remain roughly in place radially.
Angular momentum relaxation may also be ignored for our purpose
\citep[e.g.,][]{1996bRauch}.
However, the stellar distribution in the inner region that initially formed on a disc
probably has vertically thickened substantially
and we will assume that they no longer substantially contribute to local gravity 
on the gas disc (within the thickness of the assumed thin gas disc)
subsequently formed from returned stellar gas. 
Because stars in the inner regions are already
mostly rotationally supported, the shedded gas rains almost ``straight down" to 
land at a location that their specific angular momentum allows, to form a disc.
Obviously, going out radially, the rotational support lessens and star formation
may occur in a 3-d fashion.
But that does not alter our argument about what happened at small radii.
The orientation of the disc is approximately the same as the previous disc out of which stars in the inner regions
were formed, 
since the overall angular momentum distribution of stars 
has not much changed in the absence of any subsequent intrusions.
The most important difference of this new accretion disc, compared to the disc
formed during the starburst phase,
is that this new disc starts with almost no material and surface density
increases with time gradually on the timescale of hundreds of megayears to gigayear.

To have a better gauge how the results obtained depend on 
the assumed inner density slope, instead of assuming a Mestel disc as is done
in \S 3 here we present a more general case assuming 
the inner density profile of the form 
\begin{equation}
\label{eq:SigmaN}
\Sigma_g(r)= \Sigma_0 \left({r\over r_0}\right)^{-n}, 
\end{equation}
\noindent
where $n\sim [0.5,1]$ \citep[e.g.,][]{1997Faber,2009Kormendy}.
For this case Equation (\ref{eq:rQ}) is modified, taking into account 
the gradual change of the gas disc surface density with time, to be
\begin{eqnarray}
\label{eq:rQN}
r_Q &=& {1\over (\pi (3\pi)^{1/2})^{4/3(3-2n)}} \left({\dot M\over M}\right)^{2/3(3-2n)} \nonumber \\
&& (f_{\rm rec}f_g)^{-2/(3-2n)} \alpha^{-2/3(3-2n)} G^{-1/3(3-2n)} M^{5/3(3-2n)} \Sigma_0^{-2/(3-2n)} r_0^{-2n/(3-2n)}
\end{eqnarray}
\noindent
where $f_{\rm rec}$ is the total fractional stellar mass that recycles back to ISM
and 
$f_{g}(t)$ the fraction of recycled gas that has returned by time $t$ (out of the fraction $f_{\rm rec}$).
The process of SMBH accretion in this case goes as follows.
The SMBH will accrete all the gas within its Bondi radius $r_B$ over some period of time,
as long as $r_Q\ge r_B$,
where $r_B$ is defined as
\begin{equation}
\label{eq:rB}
r_B \equiv {G\mbh\over \sigma_n^2},
\end{equation}
\noindent
with $\sigma_n$ being the velocity dispersion of the inner region of the bulge
($r\le 20$pc or so for $\mbh=10^8\msun$).
For the moment we ignore any feedback effect from the SMBH.
Since $r_B$ grows with time and $r_Q$ decreases with time with increasing $f_{g}$ for $r>r_B$
that has been accumulating gas,
the condition $r_Q\ge r_B$ may be violated at some time $t$,
at which point the SMBH is cut off gas supply at its Bondi radius and
the SMBH will subsequently grow by consuming the final patch of gas on the disc
within its Bondi radius. 
Before the condition $r_Q\ge r_B$ is reached, 
the recycled gas that has landed outside (time varying) $r_B$ continues to accumulate 
(some of the accumulated gas possibly forms stars).
Using Equations (\ref{eq:rQN},\ref{eq:rB}) we find the turning point $r_Q=r_B$ is reached when
\begin{equation}
\label{eq:fg}
f_{\rm rec}f_g = {2-n\over 2}
\end{equation}
\noindent
with the disc mass within $r_Q=r_B$, i.e., SMBH mass, being
\begin{equation}
\label{eq:M8}
M_F = {3(2-n)^3\over 8} \left({\dot M\over M}\right)^{-1} {\alpha\sigma_n^3\over G}.
\end{equation}
\noindent
From Equation (\ref{eq:fg}) we see that 
$(2-n)/(2f_{\rm rec}) >1$ for $n=[0.5-1]$.
Thus, we simply correct Equation (\ref{eq:M8}) by a factor of $2f_{\rm rec}/(2-n)$ to finally arrive at
\begin{eqnarray}
\label{eq:M82}
\mbh &=& {3(2-n)^2\over 4} \left({\dot M\over M}\right)^{-1} {f_{\rm rec}\alpha\sigma_n^3\over G} \nonumber \\
 &=& 1.9\times 10^8 (2-n)^2 \alpha_{0.01} l_E^{-1} \epsilon_{0.1} \left({\sigma_n\over 200\kms}\right)^3 \msun,
\end{eqnarray}
\noindent
with the radius when $r_Q=r_B=r_{BQ}$ being:
\begin{eqnarray}
\label{eq:rBQ}
r_{BQ} = 34 (2-n)^3 \alpha_{0.01} l_E^{-1} \epsilon_{0.1}\left({\sigma_n\over 200\kms}\right){\rm pc}.
\end{eqnarray}
\noindent
Equations (\ref{eq:M82}, \ref{eq:rBQ}) suggest
that the SMBH accreted the recycled gas at $r\le 20(\mbh/10^8\msun)$~pc or so for $l_E\sim 1$;
it could be substantially larger for smaller $l_E$.
The reason that the accretable mass is so much larger during this period than
the starburst phase is because the accretion disc in this period is replenished continuously
at a moderate rate such that it is stable within a much large radius than the case of sturburst phase with a much thicker 
(surface density-wise) disk.
Equation (\ref{eq:M82}) resembles the observed $\mbh-\sigma$ relation \citep[][]{2002Tremaine}.
We argue the resemblance is deceptive, in a general sense,
because it hinges on a value of $\alpha\sim 0.01$ or so and $l_E\sim 1$.
As we mentioned earlier, the currently allowed value of $\alpha$ could range from $10^{-4}$ to $1$
and at the moment we do not know what value nature has picked to grow her SMBHs.
In light of this situation, using Equation (\ref{eq:M82}) to declare victory is premature.
However, Equation (\ref{eq:M82}) does suggest that
there is enough material and time to grow the SMBH to the observed value
during the post-starburst phase.
This is in stark contrast with the starburst phase when there is
not enough accretable matter even if one pushes the viscosity value
to the limit (see Equation \ref{eq:MQ}).

\begin{figure}[ht]
\centering
\vskip -1.2in
\resizebox{5.0in}{!}{\includegraphics[angle=0]{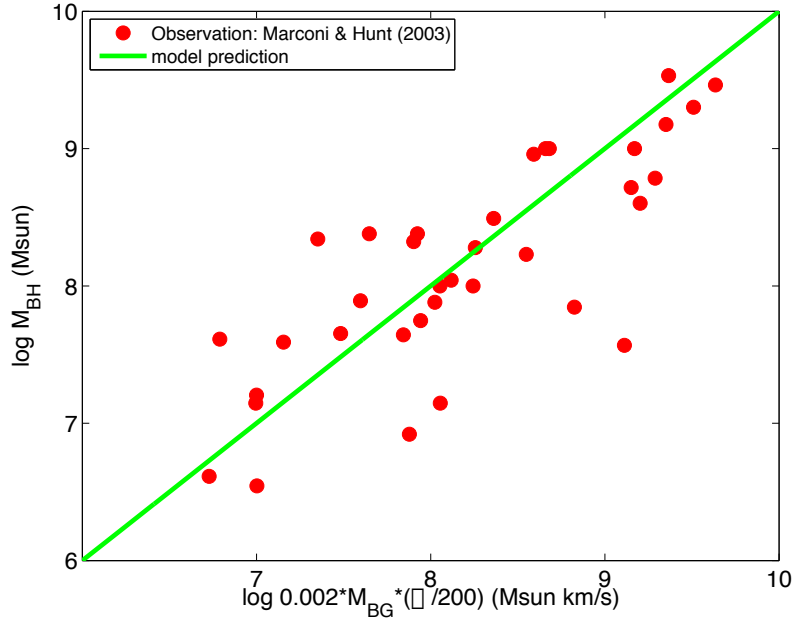}}
\vskip -1.2in
\caption{
The circles are data from \citet[][]{2003Marconi}.
The solid line is predicted by Equation \ref{eq:rat} using $A=1$ (see Equation \ref{eq:A}).
}
\label{fig:MM}
\end{figure}

A scenario where the allowed range of viscosity value is limited to one side,
i.e., $\alpha$ is allowed to have values greater than say $0.01$,
is much less fine tuned.
In this case, some self-regulation for the SMBH growth will be necessary.
This self-regulation for the SMBH growth is indeed achievable during the post-starburst phase, as we will now describe.
The total amount of radial momentum 
that radiation pressure of the SMBH may exert on the surrounding gas
is $\epsilon c \mbh$ (this is likely a lower bound, neglecting the possibility of multiple scatterings of photons
in the optically thick regime).
Equating $\epsilon \beta c\mbh$ to $f_{\rm rec}\mbg(1-f_*)v_{\rm esc}$ (that is the momentum
of the driven-way gas escaping the galaxy) gives 
\begin{eqnarray}
\label{eq:rat}
{\mbh\over \mbg} = {f_{\rm rec}v_{\rm esc}\over \epsilon\beta c} {(1-f_*)\over (1+f_{\rm rec}f_*)}
= A {2\over 1000}\sigma_{200},
\end{eqnarray}
\noindent
where $A$ is 
\begin{eqnarray}
\label{eq:A}
A = {(f_{esc}/0.15)(1-f_*)\over (1+(f_{\rm rec}/0.15)f_*)\beta\eta\epsilon_{0.1}} {v_{\rm esc}\over 2\sigma},
\end{eqnarray}
\noindent
where $f_*$ is the fraction of recycled gas that subsequently re-formed into stars and $v_{\rm esc}$ is escape velocity
[for an isothermal sphere truncated at virial radius $r_v$, $v_{\rm esc}(r)/2\sigma=(1+\ln(1+r_v/r))^{1/2}$ at 
radius $r$];
$\beta$ is the fractional solid angle that absorbs the radiation from the SMBH;
the term $(1+f_{esc}f_*)$ takes into account additional stars added to the bulge stellar mass 
formed from the recycled gas.
Of the parameters in Equation \ref{eq:A},
$f_{esc}=0.15$ is reasonable taking into account that about the half of mass return occuring at early times 
by type II supernovae can escape without additional aid;
radiative efficiency of $\epsilon=0.1\epsilon_{0.1}$ is consistent with observations
\citep[][]{2002Yu, 2004Marconi};
some fraction of the recycled gas forming into stars 
is probably unavoidable, since some gas with column density greater than Compton column
will slip through radiation pressure (see discussion below);
$f_*$ also includes the (possibly very large) amount of gas at large radii that would not have
accreted onto the SMBH in the first place even in the absence of any feedback
(e.g., molecular clouds on the Galactic disk are not being fed to the Galactic center SMBH
in a consistent fashion);
the factor $\eta$ (greater than one) takes into account additional stars that 
are not formed from the starburst event. 
Overall, considering all these balancing factors,
a value of $A$ of order unity seems quite plausible.
Figure \ref{fig:MM} plots the relation between $\mbh$ and $\sigma\mbg$ 
predicted by Equation \ref{eq:rat} using $A=1$.
It is clear that it provides a very good fit to the observed data.
A similar scaling relation as Equation \ref{eq:rat} was derived
based on a different, radio jet feedback mechanism \citep[][]{2010Soker}.

A similar scenario of linear momentum feedback from AGN radiation pressure 
has been considered by \citet[][]{2010Silk} 
to possibly produce the observed $\mbh-\mbg$ relation {\it during the starburst phase} 
but they conclude that the radiation pressure is insufficient by an order of magnitude to be 
able to blow the unwanted gas away.
The magnitude of the radiation pressure and escape velocity requirement
considered here are the same as theirs. The difference is that here
the amount of gas that need to be regulated in the post-starburst phase
is nearly a factor of $10$ lower and further allowance for star formation
from the recycled gas make possible that the radiation pressure from the central AGN
may be adequate to self-regulate the SMBH growth so as not to overgrow it.

We note that Equation \ref{eq:rat} would work without much variation
if the gas that is blown away is uniformly distributed.
The recycled gas is expected to be non-uniform. 
Even if it were uniform initially, thermal instabilities likely make the distribution non-uniform.
Given that, we elaborate further on Equations (\ref{eq:rat},\ref{eq:A}) and
the physical processes of radiation pressure driven winds.
Some distinction may be made between about 1/3 of the total solid angle
where UV and other photons are directly seen from AGN 
and the other $\beta\sim 2/3$ of the solid angle 
that has a nearly Compton thick or thicker obscuring screen, most of which probably stems
from the so-called molecular torus \citep[e.g.,][]{1999Risaliti}.
For every $\Delta M_{\rm acc}$ of mass accreted, roughly 
$\epsilon c/v_{\rm esc} \Delta M_{acc}=100 \epsilon_{0.1} (v_{esc}/300\kms)^{-1}\Delta M_{acc}$ 
of mass that rain down by aging stars could be driven away by the radiation momentum from the AGN.
In the $1/3$ opening solid angle
some portion of the radiation pressure driven winds will be accelerated to high velocities,
perhaps in a fashion similar to what is seen in simulations \citep[e.g.,][]{2009Kurosawa},
observationally manifested as broad emission or absorption lines as well as 
outflows seen in narrow lines \citep[e.g.,][]{2003Crenshaw,2011Greene}.

A significant fraction of the material may be accumulated 
in the remaining $\beta\sim 2/3$ of the solid angle (i.e., Type 2 AGNs),
including recycled gas that comes from the other 1/3 solid angle that is too heavy to
be accelerated away ``on the fly" by the radiation pressure.
In this 2/3 of the solid angle, high velocity winds radially exterior to the molecular torus 
is unlikely given the heaviness (i.e., low opacity) of the molecular torus.
We discuss some of the physics here.
To gain a more quantitative understanding, a look at some observed properties of the torus is instructive.
\citet[][]{2004Jaffe} measured the radius and height
of the molecular torus of NGC 1068 to be $1.7$pc and $2.1$pc, respectively.
The mass of the SMBH in NGC 1068
is $(8.3\pm 0.3)\times 10^6\msun$ \citep[e.g.,][]{2003Marconi}.
If we extrapolate to a $10^8\msun$ SMBH assuming that the location and height
of the molecular torus is proportional to the SMBH mass,
we have a surface area of the torus equal to $3200~$pc$^2$ at a SMBH-centric radius of $20$pc.
If we assume that the column density of the molecular torus is $10^{24}$cm$^{-2}$
\citep[e.g.,][]{1999Risaliti}, its total mass is then $2\times 10^7\msun$.
The dynamical time at $20~$pc is $10^5~$yrs.
A SMBH of mass $10^8\msun$ accreting at Eddington rate would grow
a mass of $\sim 10^5\msun$ in $10^5~$yrs, while the overall rate of gas return would be
$\sim 2\times 10^7\msun$ over the entire bulge during that period.
Thus, the abundant gas supply rate suggests that
the necessary (not sufficient) condition for a near ``steady" state is met such that 
the molecular torus may be kept roughly invariant with time, with the rate 
of driven-away gas by radiation pressure plus that
of gas forming into stars equal to the rate of gas return from aging stars.

Given the short star-formation timescale of the very dense gas in the molecular torus,
it would be unavoidable that star formation should occur there (as well as some regions exterior to it).
This ``lightens up" the torus to the extent that it may be pushed away by the radiation pressure,
when the condition that the deposited radiation momentum divided by the accumulated mass 
exceeds the escape velocity (assuming, in the absence of radiation pressure, the torus
would just be in a bound circular orbit).
In this sense the radiation momentum from the SMBH serves to retard
gas supply to accretion from the torus to let SF take over to have it mostly depleted.
In combination with the analysis in the preceding paragraph,
it seems physically plausible that radiation pressure and depletion of gas by star formation
is able to jointly reduce and regulate the amount of gas that feeds the central SMBH.
Given that the overall margin, in an ``on average" sense,
is quite thin (i.e., $A\sim 1$ in Equation \ref{eq:A}),
it is likely that there are significant variations in $A$, perhaps up to a factor of a few.

In the 1-d simulations of \citet[][]{2007Ciotti} for
an elliptical galaxy, the SMBH growth appear to be intermittent.
The intermittency in their simulations was caused by a hot X-ray heated bubble that 
prevents continued gas accretion, until it bursts, which is then followed by another
accretion episode, and so on.
We suggest that Rayleigh-Taylor instability on the shell enclosing the X-ray bubble
may prevent the X-ray bubble from inflating, 
as hinted by recent 2-d simulations of \citet[][]{2010Novak}.
It it reasonable to assume that shell fragmentation in three-dimension is still more pronounced to allow
continued deflation of a notional X-ray bubble. 
Observationally, the lack of significant X-ray emission from circumnuclear region in powerlaw elliptical galaxies
host AGNs, which we argue are the post-starburst galaxies we consider here,
supports the picture that the hot bubble is not robust \citep[e.g.,][]{2005Pellegrini}.
In the absence of a hot X-ray bubble guarding the SMBH,
we suggest that the recycled gas from aging stars is able to reach the disc 
and the accretion, with self-regulation argued above, is quasi-steady without major flares of magnitude
seen in 1-d simulations.
As we will show later, a steady declining accretion rate proportional to the 
gas return rate provides a much better match to at least two observations:
(1) the observed early-type host galaxies of AGNs 
are mostly in the green valley of the galaxy color-luminosity diagram 
with a small fraction in the red sequence (\S 5.2) \citep[e.g.,][]{2007Salim,2008Silverman,2009Hickox,2010Schawinski},
but very few in the blue cloud, which would have been the case if AGN flares are accompanied by starbursts \citep[][]{2007Ciotti};
(2) the observed AGN accretion rate for early-type galaxies in the local universe
displays a powerlaw distribution with the amplitude and decay rate 
\citep[][]{2009Kauffmann} that is  expected from the non-flare scenario that is proposed here.
This indicates that bursty AGN accretion, while quite possible and
sometimes perhaps unavoidable, is probably not the dominant mode.
It is currently a challenge but will be of great value to carry out 3-d high-resolution
simulations to more accurately quantify this outcome.

\section{Model Predictions and Discussion}

We have presented a physically motivated picture for the coevolution of 
galaxies and SMBH starting with a triggered starburst.
Let us now summarize the entire evolution in \S 5.1 and then give an incomplete list 
of implications and predictions in \S 6.2-6.9 to be qualitatively compared/verified with observations.

\subsection{Three Distinct Periods of Coevolution of Galaxies and SMBH}

\begin{figure}[ht]
\centering
\vskip -0.0in
\resizebox{6.0in}{!}{\includegraphics[angle=0]{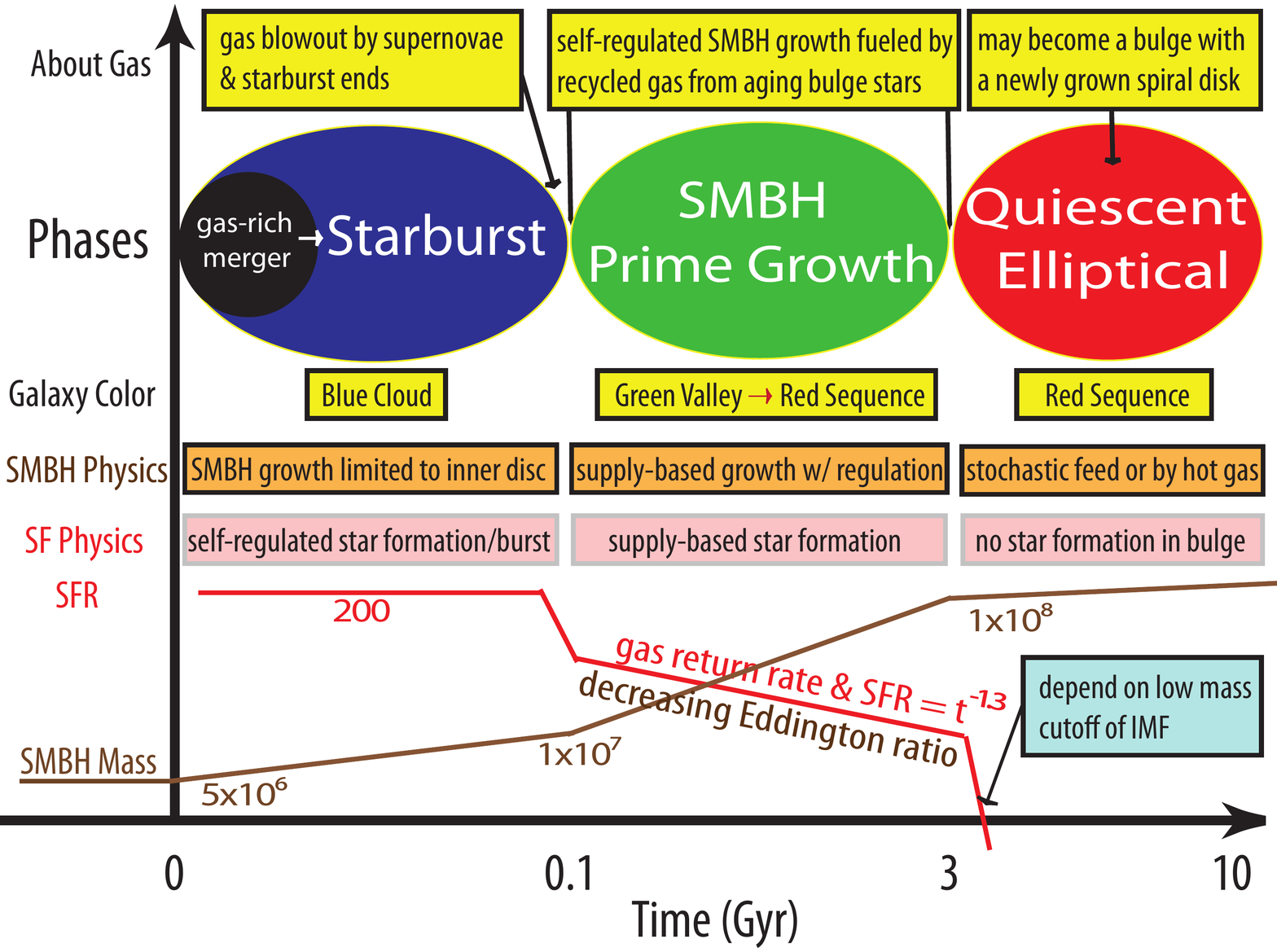}}
\vskip -0.0in
\caption{
shows the entire evolutionary process 
for an example merger of two gas-rich spirals of mass $\sim 10^{12}\msun$ each that
eventually produces a powerlaw elliptical galaxy of velocity dispersion of $200\kms$.
This scenario is not limitd to merger events but encompasses any significant
event triggering a starburst.
Note that the time boundaries between difference consecutive phases are approximate and uncertain
to within a factor of a few. 
The numbers in brown indicate the BH masses and the numbers in red indicate SFR.
These numbers are very approximate and given mainly for illustration purpose.  
Clearly, given the complexity, one should expect large variations from case to case.
}
\label{fig:evol}
\end{figure}

From the onset of a significant central starburst to becoming a quiescent bulge 
there are three distinct periods, as summarized in Figure \ref{fig:evol}
for an example merger of two gas-rich spirals each of mass $\sim 10^{12}\msun$ that
eventually becomes a powerlaw elliptical galaxy of velocity dispersion of $200\kms$.
We stress that the trigger event is not limited to major mergers.
This three-stage scenario is not new and its successes with respect many observations
have been discussed previously \citep[e.g.,][]{2004Granato, 2006Granato, 2005Cirasuolo, 2006Lapi, 2010Lamastra}.
The new theoretical element here is the primary growth of SMBH in the post 
starburst phase, which is reflected in the color and other properies of AGN hosts 
and we will show is in remarkable accord with latest observations, in contrast to the 
conventional scenario where SMBH growth primarily occurs during the starburst phase.
The time boundaries between difference consecutive phases (three ovals) are 
approximate (uncertain to a factor of at least a few). 
Given the complexity and variety of starburst trigger events, one should expect significant variations from case to case.
The expected consequences or predictions of this model 
are in many ways different from and often opposite to those of 
models that invoke AGN feedback to shut down both starburst and AGN activities
\citep[e.g,][]{1998Silk, 2006Hopkins}.
A new and in some way perhaps the most fundamental finding of this work is that
the SMBH does not grow during the starburst phase as much as previously thought, required in AGN-feedback based models,
despite the obvious condition that there is a lot of gas being ``jammed" into the central region;
this is different from almost all previous work
\citep[e.g,][]{1998Silk, 2006Hopkins,2010Debuhr} that either need to advocate very strong SMBH feedback or 
appear to overgrow the SMBH.

The idea of feeding the SMBH with recycled stellar material in the post-starburst phase 
is not new \citep[e.g.,][]{1988Norman, 2007Ciotti} and
we inherit most of the already known elements from prior work, including gas return rate
and the likelihood of continued star formation. 
Our analysis shows the likelihood that the SMBH may be 
fed too much in the post-starburst period in the absence of feedback from the SMBH,
in dramatic contrast with the starburst phase when SMBH feedback is insufficient.
While energy feedback from the SMBH certainly plays a role,
we show that the more robust 
momentum feedback from SMBH radiation pressure can play a critical role in regulating SMBH growth,
not necessarily only by blowing powerful winds,
but rather, in combination, by also pushing away thus retarding accretion of unwanted (by SMBH) gas to 
be instead consumed by star formation.
While our analysis may have captured some of the essential physics in terms of accretion and star formation demarcation,
to more realistically model the complex accretion and star formation dynamics,
much higher resolution 3-d radiation hydrodynamic simulations 
will be required and will be of tremendous value.

The ``size" of the starburst depends on the ``size" of the triggering event,
with at least some fraction of ULIRGs and SMGs due to major mergers of massive gas-rich gas.
However, irrespective of the size of the starburst event,
the time scales involved, being largely due to physics of stellar interior and
accretion time scale, remain the same.

\noindent (1) ``Starburst Period": this phase is triggered by some event.
The SMBH grows modestly during this period to possibly 
attain a mass that is up to order ten percent of its final mass.
This phase lasts about $10^7-10^8$yrs for typical starbursts and 
the host galaxies during this phase are in the blue cloud in the luminosity-color diagram.
The feedback energy/momentum from the starburst, i.e., supernovae,
drives the last patch of gas away and shuts down star formation, if needed.
In other words, the starburst is self-regulated, not by the central AGN during this period.

\noindent (2) ``SMBH Prime Period": several hundred million years 
after the end of the starburst, aging low-to-intermediate mass stars,
now in their post-main-sequence phases,
start to return a substantial fraction of their stellar mass to the ISM.
The SMBH accretion is fueled by this recycled gas lasting for order of gigayear.
The growth of SMBH is self-regulated, readily provided by the radiation pressure from the AGN.
The host galaxies during this period start out light-blue or in the ``green valley" and 
migrate to the ``red sequence".
Because the rate of gas return from stars diminishes with time and SMBH mass grows,
the Eddington ratio of the SMBH decreases with time.
The SMBH growth is synchronous with star formation from recycled gas during this period.
The accompanying star formation rate may also be substantial but typically 
does not constitute a starburst during this period. 
The entire duration of this phase depends sensitively 
on the lower cutoff mass of the initial mass function (IMF) -- a sensitive and powerful 
prediction of this model.

\noindent (3) ``Quiescent Bulge": several gigayears after the end of the starburst
the bulge is now truly red and dead - gas return rate is now negligible
so both accretion to the central SMBH and residual star formation have ceased.
It is possible that a disk is grown later around the bulge.
The feeding of the central SMBH in the bulge of spiral galaxy during this period 
is no longer by overhead material from aging stars,
rather by occasional objects that happen to be on
some plunging orbits to be disrupted by the SMBH and form a
short-lived accretion disc. 
Candidate objects may include molecular clouds, some tidally disruptable stars or gas streams.
Significant disturbances or torques, such as minor mergers and galactic bars,
could provide the necessary drivers for some more consistent accretion events.
How is a red and dead bulge with a hot atmosphere able to remain star-formation-free?
This is a major topic on its own right and beyond the scope of the current paper,
but will be addressed in a future paper.

\subsection{Some ``Obvious" Implications of the Model}

There are some unambiguous discriminating signatures of this model 
that already can be directly ``read off" Figure \ref{fig:evol}.  We highlight several here.

(1) 
Starburst and AGN growth are {\it not coeval} in this model.
AGN {\it does not} regulate the starburst, consistent with observations \citep[e.g.,][]{2009bSchawinski,2009Kaviraj}.
AGN activities is expected to outlive the starburst, in agreement with observations \citep[e.g.,][]{2008Georgakakis}.
These predictions are opposite to those of models that invoke 
AGN feedback as the primary regulating agent. 

(2) The apparent requirement of a rapid migration of early-type galaxies
from the blue cloud to the red sequence,
in order to produce a bimodal distribution in color \citep[e.g.,][]{2003bBlanton},
is primarily due to the prompt shutdown of SF by stars (i.e., supernovae) 
at the end of the starburst phase; there is no need to invoke
other ingredients, consistent with observations \citep[e.g.,][]{2010Kaviraj}.
Observationally, there is no evidence that the presence of an AGN is related to 
quenching of star formation or the color transformation of galaxies \citep[e.g.,][]{2012Aird}.
This prediction is different from that of models that invoke 
AGN feedback to quench star formation. 

(3) AGN activities in ongoing starburst galaxies, i.e., buried AGN activities,
are not expected to be dominant in this model, in agreement with observations
\citep[e.g.,][]{1998Genzel,2000Ivison,2003Ptak,2004Ivison, 2005Alexander,2005bAlexander,2006Schweitzer,
2006Kawakatu, 2008Alexander,2009Veilleux}.
Note that the above statement is not inconsistent with AGN/QSOs being
associated with galaxies in the process of merging, which may 
enhance accretion activities in the involved (yet to merge) galaxies
\citep[e.g,][]{1997Bahcall,2010Hennawi,2010Smith}.

(4) The most luminous quasars that accrete with high Eddington ratios
occur order of $100$Myr after the end of the starburst. 
They may contain substantially more merger signatures, 
which appears to be indicated by observations \citep[e.g.,][]{2008Bennert}.
If one were to identify a population in-between ULIRGs and more regular QSO hosts in terms of 
spectral properties,
they should show some more signs of tidal interactions that are yet to fully settle 
since the starburst, also consistent with observations \citep[e.g.,][]{2001Canalizo}.

(5) Low Eddington ratio AGNs that are expected to last order of Gyr
are not expected to show a close linkage to major disturbances that trigger the starburst (e.g., mergers),
since possible signatures of the trigger merger event 
have largely been erased over time,
consistent with observations \citep[e.g.,][]{2005Grogin, 2011Cisternas}.
Thus, one does not expect to see merger signatures to be associated with  
moderate-luminosity AGNs, which is in contrast with  
AGN feedback based models where most of the moderate luminosity AGNs 
are expected to coincide with starburst. 

(6) While the green-valley morphologically early-type galaxies that host AGN  
is the evolutionary link between starburst galaxies (in the blue cloud) and the red elliptical galaxies
(on the red sequence),
it is useful to distinguish between them and the other class of
green galaxies that simply continuously form a modest amount of stars (such as our own Galaxy).
The former are chronologically immediate successors to starburst galaxies
and should be in early-type galaxies,
strongly supported by observations \citep[e.g.,][]{2007Salim,2008Silverman,2009Hickox,2010Schawinski},
whereas the latter are not a chronologically intermediate class
between the blue cloud and the red sequence.
The total green galaxy population will be the sum of these two different morphological types,
with some obvious implications, such as green galaxies having mixed morphological types with limited
merger signatures, consistent with observations \citep[e.g.,][]{2011Mendez}.
This prediction is in contrast with  
AGN feedback based models where most AGN hosts 
are expected to coincide with starburst and a small fraction, mostly the most luminous AGNs 
(occuring near the end of the starburst phase), is expected
to have matured early-type morphologies. 

(7) While the early-type AGN host galaxies may have 
similar morphologies as and will eventually evolve to inactive 
elliptical galaxies, the former should have much bluer colors
than the latter, consistent with observations \citep[e.g.,][]{2004Sanchez}.
The basic morphological properties of the host galaxies of the most luminous quasars, 
corresponding to the most massive SMBHs in the prime growth phase 
should resemble those of giant elliptical galaxies,
consistent with observations \citep[e.g.,][]{2003Dunlop}.

(8) Because of the expected rate of gas return ($\propto t^{-1.3}$ on gigayear scales)
to which both SMBH accretion and star formation are proportional 
and because more powerful AGN accretion occurs closer in time to the preceding starburst,
it is expected that more powerful AGNs are hosted by early-type galaxies
with younger mean stellar ages, consistent with observations \citep[e.g.,][]{2003Kauffmann,2004Jahnke}.

(9) The accompanying star formation rate of elliptical galaxies may be 
quite substantial, on the order of $\sim (5-10) (M_*/10^{11}\msun)(t/1{\rm Gyr})^{-1.3}\msun$~yr$^{-1}$.
Thus, while most AGN host galaxies
have left the blue cloud,
a significant fraction of them, especially those hosting luminous AGNs, 
should still have substantial SFR, 
consistent with observations \citep[e.g.,][]{2009Silverman,2009Shi}.
It is expected that the incidence of star formation signatures (e.g., dust)
in the nuclear region should correlate positively with AGN activities for elliptical galaxies,
because the strengths of both are proportional to the gas return rate,
consistent with observations \citep[e.g.,][]{2007SimoesLopes}.
These predictions are opposite to AGN feedback based models where star formation
is expected to be completely quenched after AGN feedback clears the gas out.  

\subsection{Origin of Two AGN Accretion Regimes}

\citet[][]{2009Kauffmann} 
presented an insightful observational result of
two distinct regimes of black hole growth in nearby galaxies along with
its apparent implications.
They find that star-forming galaxies display a lognormal distribution of 
Eddington ratios; their interpretation is that in this regime 
accretion on to the SMBH is not limited by the supply of gas but by feedback
processes that are intrinsic to the SMBH itself.
Our model provides the following alternative interpretation for this phenomenon:
this lognormal distribution merely reflects two random processes at work:
(1) the amount of gas that landed on the stable accretion disc 
to provide accretion to the SMBH during the starburst phase
depends on many ``random" variables of the triggering event 
(in the case of a merger, such as merging orbit inclination, velocity, spin alignment, etc),
and (2) observations catch a random moment during the accretion of this gas.
Central theorem should then give rise to a lognormal distribution.
Another class of possible triggering events for SMBH accretion in star-forming galaxies
(e.g., dormant SMBH in the bulge of disk galaxies)
is stochastic feeding due to some random events, which should also follow a lognormal distribution.

Separately, they find that galaxies with old stellar populations  
is characterized by a power-law distribution function of Eddington ratios
and the AGN accretion rate is about $0.3-1$\% of the gas return
gas from recycling.
In our model the expect accretion rate is expected to be
$\mbh/(f_{\rm rec}\mbg) = 1.3\times 10^{-2} A \epsilon_{0.1}^{-1}\sigma_{200}$.
This expected relation between SMBH accretion rate and gas return rate
is remarkably close to their observed value. 
As \citet[][]{2009Kauffmann} already pointed out, the powerlaw distribution
is consistent with the recycling gas return rate $\propto t^{-1.3}$ \citep[][]{1989Mathews}.
This is a strong support for the proposed model here.

\begin{figure}[ht]
\centering
\vskip -0.3in
\resizebox{5.5in}{!}{\includegraphics[angle=0]{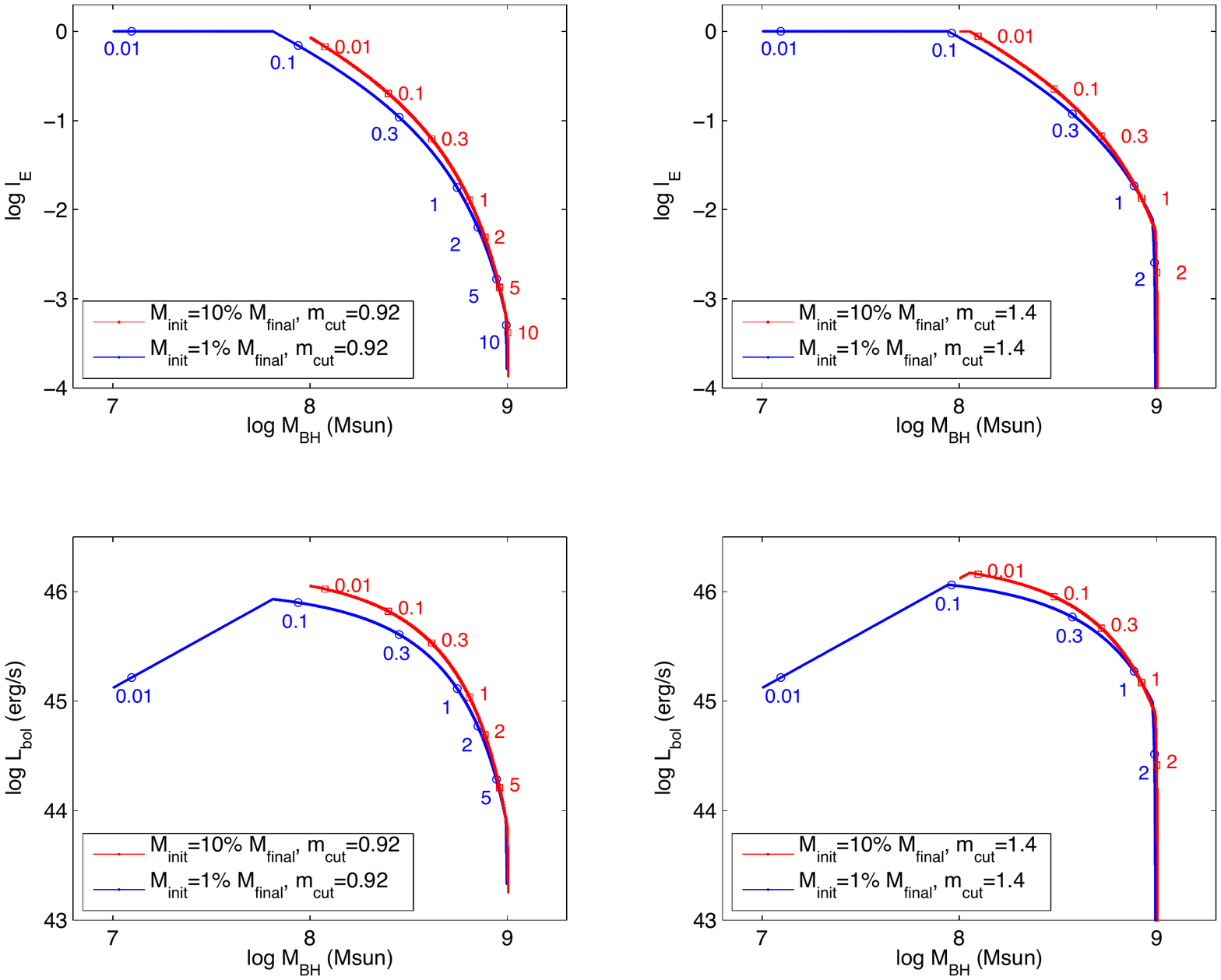}}
\vskip -0.3in
\caption{
Top left panel: evolutionary growth tracks in the SMBH mass-
Eddington ratio plane of an example SMBH 
of final mass $10^9\msun$ with two cases of seed black mass
of $10^7$ and $10^8\msun$, respectively.
A low mass cutoff for the IMF of $0.92\msun$ that
has a turnoff lifetime of $10~$Gyr is assumed.
We assume that the SMBH accretion rate is proportional 
to the recycle gas return rate of the form $\propto t^{-1.3}$ \citet[][]{1991Ciotti}
capped at the Eddington rate with a radiative efficiency
of $\epsilon=0.1$, starting $200$Myrs after the end of the starburst.
Also indicated along each track are the times in Gyrs elapsed since the start 
of the accretion.
Top right panel: the case for a low mass cutoff for the IMF of $1.4\msun$ that
has a turnoff lifetime of $2~$Gyr.
Bottom panels: tracks for the cases in top panels 
but in the SMBH mass-luminosity plane.
}
\label{fig:Mdot}
\end{figure}

\subsection{Initial Mass Function and AGN Accretion History}

Because the least massive stars live the longest,
the cutoff mass of the initial stellar mass function (IMF)
plays an important role in shaping the evolution 
on longer time scales of $\ge 1$Gyr.
For example, a $0.92\msun$ star (solar metallicity) has a lifetime of $10$Gyr,
whereas a $1.4\msun$ only lives $\sim 2$Gyr.
Thus, the duration of the ``SMBH Prime Period"
depends sensitively on the lower mass cutoff of the IMF.
Figure \ref{fig:Mdot} shows several cases of the evolution of the SMBH growth tracks.
It shows that the evolution and duration of SMBH growth in the post-starburst phase
depend sensitively on the low mass cutoff of the IMF.
We see that for a cutoff mass of $0.92\msun$ the SMBH spends about
$100$Myr accreting at Eddington limit when its mass is up to about 10\% of its final mass
and a significant period ($\ge 1$Gyr) at less than 1\% of the Eddington rate,
and most of the time at about 0.1\% of the Eddington rate
when its mass approaches its final mass.
On the other hand, with a mass cutoff of $1.4\msun$ the entire SMBH accretion shortens to $2$Gyr
and does not extend below $10^{-2}$ Eddington rate.
Since not all elliptical galaxies at present time 
are observed to accrete at $0.1\%$ of the Eddington rate,
this already suggests that a higher than $0.92\msun$ cutoff mass in the IMF
may be required.
Presently there is circumstantial evidence for massive star formation
in galactic centers, including our own Galaxy \citep[e.g.,][]{2009Lu} and 
M31 \citep[e.g.,][]{2005Bender}. 
Given the very sensitive dependence of stellar lifetime on stellar mass,
careful considerations along this line may prove to be very powerful in placing
constraints on the low-mass cutoff in the IMF as well as testing this model.
Detailed comparisons between theoretical prediction with observational data
in terms of the AGN luminosity-mass plane 
\citep[e.g.,][]{2010Steinhardt}, the Eddington ratio range \citep[e.g.,][]{2002Woo},
AGN ages at different redshifts \citep[e.g.,][]{2004Martini} 
or at different luminosities \citep[e.g.,][]{2005Adelberger} 
should also prove very powerful in constraining the IMF.
We note that our assumption used to derive the light curves 
in Figure \ref{fig:Mdot} is extremely simplistic
and therefore we do not expect that they provide satisfactory matches to observations.
It is called for that additional 
ingredients be included to account for, e.g., 
variations in stellar distribution, possible variations of IMF 
as a function of local star formation conditions, dependence of initial seed SMBH mass on
galaxy model, etc,
in order to have a more encompassing analysis.
We shall carry out a more detailed analysis with additional parameters 
in a future study, especially 
when measurements of both SMBH masses and accretion rates 
become significantly more precise for a large sample of active galaxies.

\subsection{Super-Solar Metallicity of Accreting Gas}

One clear implication is that the accretion gas,
being shedded from aging stars,
should be very metal rich with supersolar metallicity,
in agreement with observations \citep[e.g.][]{1993Hamann},
especially to explain super-solar N/He ratio \citep[e.g.,][]{1999Hamann}.
This is because nitrogen is believed to be secondary nature,
where its abundanace scales quadratically with metallicity.
The recycled gas that is feeding the SMBH in our model
fits the bill most naturally. 
In addition, the metallicity of accretion gas is not expected to depend on redshift,
being intrinsic to stellar evolution,
consistent with all accreting gas being very metal rich 
at all redshifts,
including the highest redshift SDSS quasars \citep[e.g.,][]{2006bFan}.

\subsection{Relative Cosmic Evolution Between Starburst Galaxies and AGN}

Given the modest amount of time delay (several $100$Myrs) between the starburst phase
and the SMBH prime growth phase,
it is unsurprising that one should expect to see
nearly synchronous evolution between the starburst 
and SMBH growth on longer, cosmic time scales, consistent with observations
\citep[e.g.,][]{1988Boyle,2005Nandra}.

In the context of the observed cosmic downsizing phenomenon,
the downsizing of galaxies  
\citep[e.g.,][]{1996Cowie,2005Treu} should thus be closely followed
by downsizing of AGNs \citep[e.g.,][]{2005Barger,2005Hasinger}.
There is, however, a very important difference between the two classes in post peak activities,
predicted in this model.
For starburst the shutdown time scale is expected to be 
about $\sim 100$Myrs, whereas for moderate-luminosity AGNs (i.e., Eddington ratio $\sim 10^{-3}$) 
the decay time scale is of order of $\sim 1$Gyrs.
With a deep AGN survey 
that is capable of subdividing early-type galaxies in terms of their masses,
one should be able to differentiate between the downturn time 
of starburst galaxies and that of AGNs hosted by elliptical galaxies at a fixed mass.
This prediction would be a strong differentiator between this model
and AGN-based feedback models.

\subsection{AGN Broad Emission and Absorption Lines}

Some of the overhead material raining down onto SMBH accretion disc from recycled gas
from aging low-to-intermediate mass stars 
provides the material observed as broad emission lines (BEL) and 
broad absorption lines (BAL).
When some of this gas, probably in the form of some discrete clouds,
reaches the inner region of the 
the SMBH (at $r\le 10^2r_s$, where $r_s$ is Schwarzschild radius), 
the clouds will be accelerated by radiation pressure,
likely through some absorption lines,
to velocities up to $0.1c$.
These clouds will be the observed BEL and BAL.
The fact that only 15-20\% of type I AGN to have BAL may be indicative
of the discrete nature of the clouds, 
not unexpected from discrete stellar remnants or from cooling instabilities.

An advantage of this overhead material 
is that it naturally provides gas clouds that are presumably to 
be some $\ge 50^o$ off the equatorial plane, in order not to be obscured 
by the molecular torus (there are of course BEL and BAL gas clouds at smaller angles but they
are not seen directly).
In this model we do not need any additional pressure force 
to lift the gas off the accretion disc - 
some of the raining down gas clouds from aging stars will be 
launched outwards before they reach the disc, physics of which
is well known \citep[e.g.,][]{1995Murray}.

\subsection{Evolution of SMBH Mass Relative to Bulge Mass}

Massive elliptical galaxies appear to have increased their masses
by $30-100\%$ in the last $7$Gyr \citep[e.g.,][]{2008Brown}.
The growth of the elliptical mass is not expected to be 
always accompanied by corresponding growth in the mass of the central SMBH.
For example, merger of a spiral galaxy without a significant SMBH and 
an elliptical galaxy would make the final SMBH appear less massive.
Given the dependence of $\mbh/\mbg \propto \sigma \propto (1+z)^{1/2}$ predicted
in this model, we predict that the $\mbh/\mbg$ relation should evolve with redshift
stronger than $(1+z)^{1/2}$ for quiescent elliptical galaxies.

\subsection{On Relation between SMBHs and Pseudo-bulges}

It is useful to add a note on the difference between classic bulges
and pseudo-bulges \citep[][]{2004Kormendy} with respect to the central SMBHs 
in this model.
The relation derived, Equations (\ref{eq:rat}, \ref{eq:A}),
that matches the observed $\mbh-\mbg$ relation
is dependent on the abundant supply of recycled gas in the inner region.
Given the sufficient gas supply from recycled gas, the feedback from the SMBH 
then can regulate its own growth.
This essential ingredient of sufficient gas supply is consistent with the observed inner slope of
classic bulges \citep[e.g.,][]{1997Faber,2009Kormendy},
as we have shown.

The situation would be very different, if star formation is not as centrally 
concentrated as in classic bulges,
for example, in rings \citep[][and references therein]{2004Kormendy}
of high angular momentum with a hollow core.
In this case, the amount of recycled gas raining down from the innermost region
may depend on other unknown factors.
For instance, if secular processes act promptly, compared to the time scales
of stellar gas recycle ($\sim 0.1-1~$Gyr), to be able to substantially fill the central region
with stars initially formed in outer regions,
the SMBH may follow the track we described.
If, on ther other hand,
secular processes evolve on longer time scales,
the recycled stellar gas would predominantly 
land in outer regions that do not efficiently accrete to the SMBH,
which would in turn not grow substantially.
It would seem likely that there may be two trends for pseudo-bulges:
(1) there will be large variations in $\mbh-\mbg$ relation and (2) SMBH 
masses may lie below that of the $\mbh-\mbg$ relation
derived from inactive classic elliptical galaxies/bulges,
both consistent with independent considerations in the context of hierarchical structure formation model \citep[e.g.,][]{2012Shankar}.
Observations, while very challenging, may have already provided some hints of both 
\citep[][]{2008Greene}.

Moreover, we do not expect 
any discernible correlation between the SMBH and galaxy disk or dark matter halo,
simply because the stars in disks do not affect SMBH growth and
the overall dark matter halo, while indirectly affect the escape velocity
that enters Equation (\ref{eq:A}), does not control the amount of gas that feeds the SMBH.
This prediction is consistent with observations \citep[e.g.,][]{2011Kormendy}.
In addition, some stellar population in the outskirts (either on a disk or just at large radii
of an elliptical galaxy) of AGN hosts may be unrelated to the preceding starburst and could be
substantially different from bulge stars \citep[e.g.,][]{2001Nolan}.

\section{Conclusions}

We have presented an alternative physical model that has the following 
characteristics for the coevolution of galaxy and SMBH.
From the onset of a starburst to becoming a quiescent bulge (in the absence 
of any subsequent significant burst event) there are three distinct periods:

\noindent (1) ``Starburst Period": some significant event induces a starburst that probably lasts about $10^7-10^8$yrs.
The SMBH grows modestly during this period to possibly 
attain a mass that is up to order ten percent of its final mass.
The feedback energy/momentum from the starburst, i.e., supernovae,
drives the last patch of gas away and shuts down star formation.

\noindent (2) ``SMBH Prime Period": several hundred million years 
after the end of the starburst, aging low-to-intermediate mass stars,
now in their post-main-sequence phases,
start to return a substantial fraction of their stellar mass to the ISM.
Because the rate of gas return from stars diminishes with time,
the Eddington ratio of the SMBH decreases with time roughly as $t^{-1.3}$.
The SMBH growth is synchronous with star formation from recycled gas during this period.
The accompanying star formation rate may also be substantial. 
The duration of this phase depends sensitively on the lower cutoff mass of the initial mass function (IMF).

\noindent (3) ``Quiescent Bulge": on order of gigayear after the end of the starburst
the elliptical galaxy is now truly red and dead - gas return rate is now negligible
so both accretion to the central SMBH and residual star formation have ceased.
It is possible that a disk may grow around the bulge later.
The feeding of the central SMBH in the bulge of spiral galaxy during this period 
is not by overhead material from aging stars,
rather by occasional objects that happen to be on
some plunging orbits to be disrupted by the SMBH and form a
short-lived accretion disc.
Candidate objects may include molecular clouds or tidally disrupted stars.

In this model, the end of starburst precedes the onset of prime SMBH growth 
by order of $100$Myr.
Starburst is responsible for shutting down its own activities; AGN has little to do with it.
AGN does provide self-regulation during its prime growth post-starburst period.
An important feature of this model is that it is physically based
and no significant fine tuning is required.
The physical reason why the SMBH does not grow substantially
in the starburst phase, although over-supplied with gas,
is that only a very small central
disc is gravitationally stable for gas accretion onto the SMBH, 
while all other regions are unstable and more conducive to star formation.
The condition is just the opposite during the post-starburst phase where
recycled gas dropout from aging stars returns slowly and can be 
more effectively accreted, so effective that self-regulation comes to play,
energetically feasibly provided by the radiation pressure.

Many comparisons between this physical model and extant observations are made
and the model appears to be in very agreement with them, including the $\mbh-\mbg$ relation.
This model predicts that the distribution of the  
Eddington ratio of AGNs in star-forming galaxies is lognormal,
whereas that of AGNs in early type galaxies is a powerlaw,
consistent with observations \citep[][]{2009Kauffmann}.
We predict that early-type galaxy hosts of high Eddingtion rate AGNs are
expected to be light-blue to green in optical color, gradually evolving to the red sequences with decreasing AGN luminosity.

\acknowledgments

I thank Jerry Ostriker for useful comments and discussion.
I would also like to thank Greg Novak, Kevin Schawinski and Charles Steinhardt
for useful discussion. I thank an anonymous referee for critical and constructive reports.
This work is supported in part by grants NNX08AH31G and NNX11AI23G.


\end{document}